\documentclass[12pt]{article}
\usepackage[pctex32]{graphics}
\begin{document}
~
\vspace{0.5cm}
\begin{center} {\Large \bf  Spin Chain with Magnetic Field and Spinning String in Magnetic Field Background }
                                                  
\vspace{1cm}

                      Wung-Hong Huang\\
                       Department of Physics\\
                       National Cheng Kung University\\
                       Tainan, Taiwan\\

\end{center}
\vspace{1cm}
\begin{center} {\large \bf  Abstract} \end{center}
We analyze the fast-moving string in the magnetic Melvin field background and find that the associated effective Lagrangian of string sigma model describes the spin chain model with external magnetic field.   The spin vector in the spin chain has been properly deformed and is living on the deformed two-sphere or deformed two-dimensional hyperboloid, depending on the direction around which the string is spinning.  We describe in detail the characters of spin deformation and, in particular, see that this is a general property for a string moving in a class of deformed background. 
\vspace{4cm}
\begin{flushleft}
E-mail:  whhwung@mail.ncku.edu.tw\\
\end{flushleft}
\newpage
\section {Introduction}
In a recent paper [1] the continuum limit of the $SU(2)$ Heisenberg spin chain was shown to reproduce the action describing string rotating with large angular momentum in  $S^3$.  This result establishes the relation of the integrable spin chain system to the string non-linear sigma model and thus  provides a very powerful tool for analyzing the integrable structures that arise on both sides of the correspondence as well as for improving our understanding of the AdS/CFT correspondence [2].  The generalization to the compact $SU(3)$ spin chain that dual to strings rotating in  $S^5$ was worked out in [3].  The paper [4] had shown a match of the noncompact $SL(2)$ spin chain sigma model  to the  string spinning on $S_\phi^1\times S_\varphi^1$ with  $S_\phi^1\in$~$AdS_5$ and $S_\varphi^1\in S^5$, 

  The powerful method of the Bethe ansatz on the spin system encourages us to establish the correspondence between the classical string state and spin system.   Therefore it is interesting to see how the correspondence could be found in the more general background, especially, those with less supersymmetry from  the point of view of phenomena.   The interesting paper [5] had shown  a match of the anisotropic XXZ spin chain sigma model  to the  string spinning on the less supersymmetry of $\beta$-deformation Lunin-Maldacena background [6], which related to ${\cal N} =1 $ $\beta$ deformed SYM.  

  Historically, a simple mechanism to break the supersymmetry is to introduce the magnetic field as proposed by Bachas [7].  This is because that the magnetic fields couple differently to particles of different spins they naturally break supersymmetry.    Therefore it is instructive to investigate the correspondence between the spin chain and spinning string in the magnetic field background.  As the AdS/CFT and AdS/Spin-chain correspondence is established in the $AdS_5 \times S^5$ spacetime the spacetime we considered will be the magnetic field deformed  $AdS_m \times S^n$, which has been constructed by us in [8] and [9].  

In the this paper we will follow the methods of  [1,4] to analyze the spinning string in the magnetic field deformed  $AdS_5 \times S^5$.  In section II  we will see that the associated effective Lagrangian of string sigma model describes the spin chain model with magnetic field.  The spin vector in the spin chain has been properly deformed and is living on the deformed $S^2$.   Especially,  We discuss in detail the characters of spin deformation  and show that the spin deformations are the general property in a class of deformed spacetime.   In section III we investigate the same problem while with the spin vector living on the deformed two-dimensional hyperboloid.   The last section is devoted to discussion.

Note that there are papers treating string backgrounds with non-trivial magnetic field [11], however, the spacetimes do not  behave as a deformed $AdS_5 \times S^5$ it could not reproduce a corresponding spin chain model.  In the recent papers [5,12] the cases of non-trivial deformations of the $AdS_5 \times S^5$  are considered where the Bethe equations are also obtained.   

\section {Deformed $SU(2)$ Spin Chain and Spinning String in Magnetic Field Deformed background}
The relevant part of the metric in the coordinate we used is described as
$$ds^2 = -A(\psi) dt^2+ B(\psi) d\psi^2+ C(\psi) d\phi_1^2 +D(\psi) d\phi_2^2. \eqno{(2.1)}$$
For the magnetic field deformed  $AdS_5 \times S^5$ spacetime which is found in our previous paper [8] the functions in (2.1) are
$$ A(\psi) = B(\psi)= \sqrt{1+ B_0^2 \cos^2\psi},~~C(\psi)= {\cos^2\psi\over\sqrt{1+ B_0^2 \cos^2\psi}},~~D(\psi)= \sqrt{1+ B_0^2 \cos^2\psi}~\sin^2\psi, \eqno{(2.2)}$$
respectively.  To find the string sigma model of the spinning string we will following the method of [1].   First,  making a change of coordinate $\phi_1 = \varphi_1+\varphi_2$ and $\phi_2 = \varphi_1-\varphi_2$ and using a replacing $\varphi_1\rightarrow t+\varphi_1$, then the line element (2.1) becomes
$$ds^2 = -(A-C-D) dt^2+ B d\psi^2+ (C+D) (d\varphi_1^2 + d\varphi_2^2) + 2(C-D) d\varphi_1 d\varphi_2\hspace{2cm}$$
$$ \hspace{5cm}+2(C+D) dt d\varphi_1 + 2(C-D) dt d\varphi_2. \eqno{(2.3)}$$
In considering a rapidly spinning string we can let $t \rightarrow \kappa \tau$ in which
$$\kappa \rightarrow \infty, ~~~~\partial_t X^\mu \rightarrow 0,~~~~~with~\kappa \partial_t X^\mu \rightarrow ~finite.\eqno{(2.4)}$$
Then the associated Lagrangian could be approximated as a simple form 
$${\cal L} \equiv G_{\mu\nu} \partial_t X^\mu \partial_t X^\nu - G_{\mu\nu} \partial_\sigma X^\mu \partial_\sigma X^{\nu} \approx  -(A-C-D) - B (\partial_\sigma\psi)^2\hspace{3cm}$$
$$ \hspace{3cm}- (C+D) ((\partial_\sigma\varphi_1)^2+(\partial_\sigma\varphi_2)^2) -2(C-D)(\partial_\sigma\varphi_1~\partial_\sigma\varphi_2),  \eqno{(2.5)}$$
A useful Virasoro condition becomes
$$0=G_{\mu\nu} \partial_\sigma X^\mu \partial_t X^\nu \approx (C+D)~\partial_\sigma\varphi_1 +(C-D)~\partial_\sigma\varphi_2.  \eqno{(2.6)}$$
Using (2.6) the Lagrangian (2.5) becomes
$${\cal L} = -(A-C-D) - B (\partial_\sigma\psi)^2 - {4CD\over C+D}(\partial_\sigma\varphi_2)^2.  \eqno{(2.7)}$$
In (2.5) we have neglected the terms of $\dot \varphi_1$ and $\dot \varphi_2$ as they are irrelevant to the static spin chain model which we are considering for a simplicity.  Another Virasoro condition $G_{\mu\nu} \partial_t X^\mu \partial_t X^\nu + G_{\mu\nu} \partial_\sigma X^\mu \partial_\sigma X^{\nu}$ will just determine $\varphi_1$ as a function of $\sigma$ and $\tau$ and is irrelevant to our investigations, as that in [1].  Using the angular relations $\psi ={1\over 2}\theta$ and $\varphi_2 = {1\over 2}\phi$ [1] the above Lagrangian becomes
$${\cal L} = -\left[A(\theta)-C(\theta)-D(\theta)\right] - {1\over 4 }B(\theta) (\partial_\sigma\theta)^2 - {C(\theta)D(\theta)\over C(\theta)+D(\theta)}(\partial_\sigma\phi)^2.  \eqno{(2.8)}$$
Our next work is to find a spin chain Hamiltonian which is described by the string sigma model of the Lagrangian (2.8).   First, we will define the spin vector $\vec n$ as 
$$\vec n = (n_x, n_y, n_z)  = \left(\sqrt{4C(\theta)D(\theta)\over C(\theta)+D(\theta)}~\cos\phi, \sqrt{4C(\theta)D(\theta)\over C(\theta)+D(\theta)}~\sin\phi,  F(\theta)\right).\eqno{(2.9)}$$
Then, for the Hamiltonian term $\sim \partial_\sigma \vec n \cdot \partial_\sigma \vec n$ to reproduce the derivative terms of (2.8) it is easy to see that the following relation shall be satisfied.
$$\left(\partial_\theta\left(\sqrt{4C(\theta)D(\theta)\over C(\theta)+D(\theta)}\right)\right)^2+\left(\partial_\theta F(\theta)\right)^2  = {B(\theta)\over4}.\eqno{(2.10)}$$
Thus we conclude that solving (2.10) and substituting the function $F(\theta)$ into (2.9) the Hamiltonian term $\sim \partial_\sigma \vec n \cdot \partial_\sigma \vec n$ could reproduce the derivative terms of (2.8).  Note that as the first term in (2.8) does not contain derivation it could be read as an energy from the external magnetic field or from a site interaction. Also, in general, the spin vector of (2.9) could not satisfies the relation $\vec n \cdot \vec n =1$, it therefore describes a deformed spin in the general background (2.1).

However, one does not be able to solve (2.10) in the general background.   Even for the simple spacetime of (2.2) the function $F(\theta)$ could not be found exactly.   We will therefore consider the spacetime with a small Melvin $B_0$ field.  In this case the spin vector becomes
$$n_x \approx \left(1+{B_0^2\over 8}(\cos\theta+ \cos^2\theta)\right)\sin\theta\cos\phi,\eqno{(2.11a)}$$
$$n_y \approx  \left(1+{B_0^2\over 8}(\cos\theta+ \cos^2\theta)\right)\sin\theta\sin\phi,\eqno{(2.11b)}$$
$$n_z \approx \left(1+{B_0^2\over 8}(1+ \cos\theta+ \cos^2\theta)\right)\cos\theta, \eqno{(2.11c)}$$
and energy from external magnetic field is 
$$\left[A(\theta)-C(\theta)-D(\theta)\right] \sim {B_0^2}~(1+ 2\cos\theta+ \cos^2\theta) \approx   {B_0^2}~(1+ 2 n_z+ n_z^2).\eqno{(2.12)}$$
Thus the corresponding spin chain model is described by a Hamiltonian
$${\cal H} \sim \partial_\sigma \vec n \cdot \partial_\sigma \vec n + {B_0^2}~(1+ 2 n_z+ n_z^2).\eqno{(2.13)}$$
in which the deformed spin vector is described by (2.11).  Note that the term which is proportional to $n_z$ is a dipole type interaction and that is proportional to $n_z^2$ is a quadrupole type interaction. 

  The spin vector in the SU(2) theory [1] is living on $S^2$ of SU(2) while that in the SL(2) theory [4] is living on hyperboloid of SL(2).  Our spin vector described in (2.11) is living on deformed $S^2$ of SU(2) as can be seen after plotting the spin vector (2.11) in three dimensional space.    For a clear illustration we show in figure 1 the $\phi =0$ section of  deformed $S^2$.  
\\
\\
\scalebox{1}{\hspace{5cm}\includegraphics{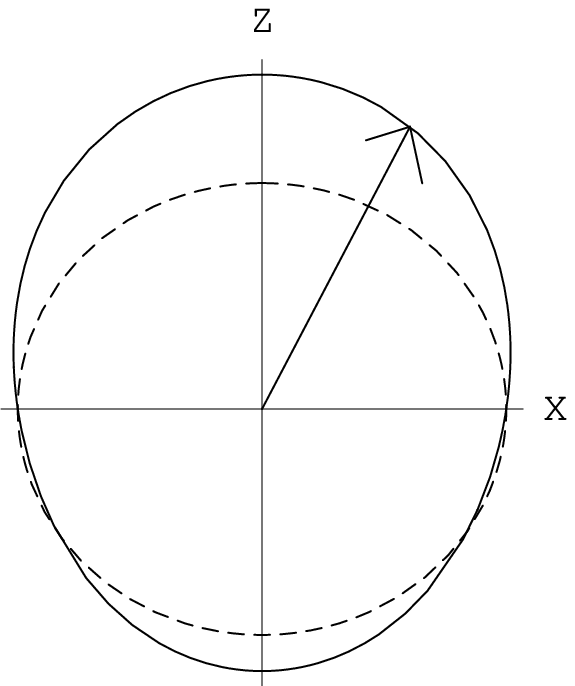}}
\\
{\hspace{3cm} {\it Figure 1. The deformed $S^2$ on which the spin is living.   The dashed line is a circle on which the spin is living for the undeformed theory.  The solid line plots the $\phi =0$ section of  deformed $S^2$ on which the spin is living.}}
\\
\\
Finally, let us consider the effect of EM vector potential $A_{\phi_1} = {B_0\cos^2\psi\over 2(1+B_0^2\cos^2\psi)}$, which is shown in the Melvin field deformed spacetimes [8].  The corresponding action is known to be
$${\cal S_A}\sim \int d\sigma \partial_\psi A_{\phi_1}\left(\partial_\sigma \psi \partial_t\phi_1 -\partial_t\psi  \partial_\sigma \phi_1\right) \approx \int d\sigma \partial_\psi A_{\phi_1}\partial_\sigma \psi= \int d\sigma  \partial_\sigma A_{\phi_1}=0,\eqno{(2.14)}$$
in which we consider the rapidly spinning string and as the string is closed, $A_{\phi_1}$ is a periodic function of $\sigma$ which implies that the above action is zero.

In conclusion, the spin chain Hamiltonian (2.13) with the deformed $SU(2)$ spin described in (2.11) corresponds to the rapidly spinning string in the magnetic Melvin field deformed background (2.2).

\section {Deformed $SL(2)$ Spin Chain and Spinning String in Magnetic Field Deformed Background}
The relevant part of the metric in the coordinate we used is described as
$$ds^2 = -A(\rho_0) dt^2+ B(\rho_0) d\rho_0^2+ C(\rho_0) d\phi_1^2 +D(\rho_0) d\varphi_3^2. \eqno{(3.1)}$$
For the magnetic field deformed  $AdS_5 \times S^4$ spacetime which is found in our previous paper [9] the functions in (2.1) are $D(\rho_0)=1$ and
$$ A(\rho_0) = \cosh^2\rho_0 \sqrt{1+ B_0^2 \sinh^2\rho_0},~~B(\rho_0) =\sqrt{1+ B_0^2 \sinh^2\rho_0},~~C(\rho_0)= {\sinh^2\rho_0\over \sqrt{1+ B_0^2 \sinh^2\rho_0}}, \eqno{(3.2)}$$
respectively.  To find the string sigma model of the spinning string we will following the method of [4].   First,  making a change of coordinate
 $$\phi_1 = \phi+t,~~~\varphi_3 = \varphi+t,~~~~\rho_0 = \rho/2, \eqno{(3.3)}$$
and in considering a rapidly spinning string we let $t \rightarrow \kappa \tau$ in which
$$\kappa \rightarrow \infty, ~~~~\partial_t X^\mu \rightarrow 0,~~~~~with~\kappa \partial_t X^\mu \rightarrow ~finite.\eqno{(3.4)}$$
Then the  associated Lagrangian could be approximated as a simple form 
$${\cal L} \approx  -(A-C-D) - {B\over 4} (\partial_\sigma\rho)^2 - C (\partial_\sigma\phi)^2 - D (\partial_\sigma\varphi)^2.  \eqno{(3.5)}$$
A useful Virasoro condition becomes
$$0 \approx  C~\partial_\sigma\phi +D\partial_\sigma\varphi.  \eqno{(3.6)}$$
Using (3.6) the Lagrangian (3.5) becomes
$${\cal L} = -(A-C-D) - {B\over 4} (\partial_\sigma\rho)^2 - \left(C+{C^2\over D}\right) (\partial_\sigma\phi)^2.  \eqno{(3.7)}$$
In (3.5) we have neglected the terms of $\dot \varphi_1$ and $\dot \varphi_2$ as they are irrelevant to the static spin chain model which we are considering for a simplicity.  As before, the another Virasoro condition will just determine $\varphi$ as a function of $\sigma$ and $\tau$ and is irrelevant to our investigations.

  Our next work is to find a spin chain Hamiltonian which is described by the string sigma model of the Lagrangian (3.7).   First, we will define the spin vector $\vec n$ as 
$$\vec n = (n_x, n_y, n_z)  = \left(\sqrt{C+{C^2\over D}}\sin\phi,  \sqrt{C+{C^2\over D}}\cos\phi , F(\rho)\right).\eqno{(3.8)}$$
Then, for the Hamiltonian term $\sim \partial_\sigma \vec n \cdot \partial_\sigma \vec n$ to reproduce the derivative terms of (3.7) it is easy to see that the following relation shall be satisfied.
$$\left(\partial_\rho\left(C(\rho)+ \sqrt{C(\rho)^2\over D(\rho)}\right)\right)^2 +\left(\partial_\rho F(\rho)\right)^2 = {B(\rho)\over 4}.\eqno{(3.9)}$$
Thus we conclude that solving (3.9) and substituting the function $F(\rho)$ into (3.8) the Hamiltonian term $\sim \partial_\sigma \vec n \cdot \partial_\sigma \vec n$ could reproduce the derivative terms of (3.7).  Note that as the first term in (3.7) does not contain derivation it could be read as an energy from the external magnetic field or from a site interaction. Also, in general, the spin vector of (3.8) could not satisfies the hyperbolic relation $ n_z^2-n_x^2-n_y^2  =1$, it therefore describes a deformed $SL(2)$ spin in the general background (3.1).

However, one does not be able to solve (3.9) in the general background.   Even for the simple spacetime of (3.2) the function $F(\rho)$ could not be found exactly.   We will therefore consider the spacetime with a small Melvin $B_0$ field.  In this case the spin vector becomes
$$n_x \approx \left(\sinh \rho - B_0^2\sinh^4{\rho\over 2}\coth\rho\right)\cos\phi, \hspace{2.5cm}\eqno{(3.10a)}$$
$$n_y \approx \left(\sinh \rho - B_0^2\sinh^4{\rho\over 2}\coth\rho\right)\sin\phi, \hspace{2.5cm}\eqno{(3.10b)}$$
$$n_z \approx  \cosh\rho+ {B_0^2\over 32}\left(-24 \ell n(1+\cosh\rho)+\left(8+7\cosh\rho\right.\right.$$
$$\left.\left.\hspace{4cm}+2\cosh 2\rho- \cosh 3\rho\right) sech^2{\rho\over 2}\right), \eqno{(3.10c)}$$
and energy from external magnetic field is 
$$\left[A(\theta)-C(\theta)-D(\theta)\right] \sim {B_0^2}~(\cosh\rho - \cosh^2\rho) \approx   {B_0^2}~( n_z- n_z^2).\eqno{(3.11)}$$
Thus the corresponding spin chain model is described by a Hamiltonian
$${\cal H} \sim \partial_\sigma \vec n \cdot \partial_\sigma \vec n + {B_0^2}~( n_z- n_z^2).\eqno{(3.12)}$$
in which the deformed spin vector is described by (3.10).  Note that the term which is proportional to $n_z$ is a dipole type interaction and that is proportional to $n_z^2$ is a quadrupole type interaction. 

  The spin vector described in (3.10) is living on the deformed  two-dimensional hyperboloid of $SL(2)$ as can be seen after plotting the spin vector (3.10) in three dimensional space.    For a clear illustration we show in figure 2 the $\phi =0$ section of  deformed $SL(2)$.  
\\
\\
\scalebox{1}{\hspace{4cm}\includegraphics{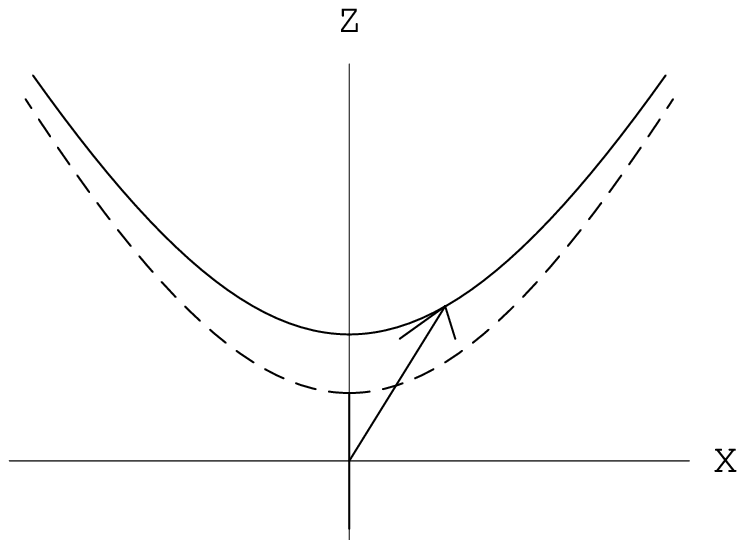}}
\\
{\hspace{3cm} {\it Figure 2. The deformed two-dimensional hyperboloid on which the spin is living.   The dashed line is an undeformed hyperboloid on which the spin is living for the undeformed theory.  The solid line plots the $\phi =0$ section of  deformed hyperboloid on which the spin is living.}}
\\
\\
Finally, let us consider the effect of EM vector potential $A_{\phi_1} = {B_0\sinh^2\rho_0\over 2(1+B_0^2\sinh^2\rho)}$, which is shown in the Melvin field deformed spacetimes [9].  Then, like those analyses in the previous section, in  considering  the rapidly spinning string as the string is closed, $A_{\phi_1}$ is a periodic function of $\sigma$ which implies that the corresponding part in the action is zero.

In conclusion, the spin chain Hamiltonian (3.12) with the deformed $SL(2)$ spin described in (3.10) corresponds to the rapidly spinning string in the magnetic Melvin field background (3.2).
  
\section{Conclusion}
The powerful method of the Bethe ansatz on the spin system encourages us to establish the correspondence between the classical string state and spin system.  
The paper [1] had shown that the continuum limit of the $SU(2)$ Heisenberg spin chain reproduces the action describing string rotating with large angular momentum in  $S^3$.   The generalization to the compact $SU(3)$ spin chain and noncompact $SL(2)$ spin chain sigma model  had also be analyzed in [3,4].

It is interesting to see how the mechanism of the correspondence could be found in the more general background, especially, those with less supersymmetry from  the point of view of phenomena.   The interesting paper [5] had showed  a match of the anisotropic XXZ spin chain sigma model  to the  string spinning on the less supersymmetry of $\beta$-deformation Lunin-Maldacena background [6], which related to ${\cal N} =1 $ $\beta$ deformed SYM.  In this paper we consider the spinning string in the less supersymmetric spacetime of the magnetic Melvin field deformed $AdS\times S$ background [8,9] to see how the  deformation will affect the AdS/spin-chain correspondence. 

We follow the analyses of  [1] and [4] to investigate the fast-moving string in the magnetic Melvin background and find that the string rotating with large angular momentum in the deformed background reproduces the spin chain model with external magnetic field.  We have shown that the $SU(2)$ or $SL(2)$ spin vector in the corresponding spin chain has been properly deformed and is living on the deformed two-sphere or deformed two-dimensional hyperboloid, depending on the direction around which the string is spinning.  We describe in detail the characters of spin deformation and have seen that the behavior of spin deformation is a general property in a class of deformed spacetime.  
\\
~
\\
~
\\
~
\\
{\Large \bf  References}
\begin{enumerate}
\item M.~Kruczenski, ``Spin chains and string theory,'' Phys. Rev. Lett. 93 (2004) 161602. h[ep-th/0311203].
\item J.~M.~Maldacena, ``The large N limit of superconformal field theories and supergravity,'' Adv.\ Theor.\ Math.\ Phys.\  {\bf 2}, 231 (1998) [Int.\ J.\ Theor.\ Phys.\  38 (1999) 1113  [hep-th/9711200]; S.~S.~Gubser, I.~R.~Klebanov and A.~M.~Polyakov, ``Gauge theory correlators from non-critical string theory,'' Phys.\ Lett.\ B428 (1998) 105 [hep-th/9802109]; E.~Witten, ``Anti-de Sitter space and holography,'' Adv.\ Theor.\ Math.\ Phys.\   2 (1998) 253 [hep-th/9802150].
\item  R.~Hernandez and E.~Lopez, ``The SU(3) spin chain sigma model and string theory,'' JHEP 0404 (2004) 052 [hep-th/0403139].
\item  S. Bellucci, P. Y. Caesteill, J. F. Morales and C. Sochichi, ``SL(2) spin chain and spinning strings on $AdS_5 \times S^5$,'' Nucl.\ Phys.\ B707 (2005) 303 [hep-th/0409086].
\item S.~A.~Frolov, R.~Roiban and A.~A.~Tseytlin, ``Gauge-string duality for superconformal deformations of N = 4 super Yang-Mills theory,'' JHEP 0507 (2005) 045 [hep-/0503192].
\item  O.~Lunin and J.~Maldacena, ``Deforming field theories with U(1) x U(1) global symmetry and their gravity duals,'' JHEP  0505  (2005)  033  [hep-th/0502086]. 
\item C. Bachas, ``A way to  Break Supersymmetry,'' [hep-th/9503030];  Micha Berkooz, Michael R. Douglas, Robert G. Leigh, ``Branes Intersecting at Angles,'' Nucl.Phys. B480 (1996) 265-278 [hep-th/9606139].
\item  Wung-Hong Huang, ``Semiclassical Rotating Strings in  Electric and Magnetic Fields Deformed $AdS_5 \times S^5$ Spacetime'', Phys. Rev. D73 (2006) 026007 [hep-th/0512117]; Wung-Hong Huang, ``Electric/Magnetic Field Deformed Giant Gravitons in Melvin Geometry'', Phys.Lett. B635 (2006) 141 [hep-th/0602019 ].  
\item  Wung-Hong Huang, ``Multi-spin String Solutions in Magnetic-flux Deformed $AdS_n \times S^m$ Spacetime'', JHEP 0512 (2005) 013 [hep-th/0510136 ].
\item C. Bachas, ``D-brane dynamics'', Phys.Lett. B374 (1996) 37 [hep-th/9511043
]; E. Alvarez, J.L.F. Barbon, J. Borlaf, ``T-duality for open strings'', Nucl.Phys. B479 (1996) 218 [hep-th/9603089];  E. Bergshoeff, M. de Roo, `` D-branes and T-duality'', Phys.Lett. B380 (1996) 265 [hep-th/9603123].
\item A.A. Tseytlin, ``Exact solutions of closed string theory'', Class.Quant.Grav. 12 (1995) 2365 [hep-th/9505052] ; ``Closed strings in uniform magnetic field backgrounds'', [hep-th/9505145].
\item S.A. Frolov, R. Roiban, A.A. Tseytlin, ``Gauge-string duality for (non)supersymmetric deformations of N=4 Super Yang-Mills theory'', Nucl.Phys. B731 (2005) 1-44 [hep-th/0507021];  T. McLoughlin, I. Swanson, ``Integrable twists in AdS/CFT'', [hep-th/0605018].
\end{enumerate}
\end{document}